\documentclass[twocolumn,showpacs,preprintnumbers,amsmath,amssymb]{revtex4}

\usepackage{graphicx} 
\usepackage{dcolumn} 
\usepackage{bm} 

\begin{document}

\title{Thermopower of multilayer graphene}

\author{Lei Hao and T. K. Lee}
\address{Institute of Physics, Academia Sinica, NanKang, Taipei 11529, Taiwan}

\date{\today}

\begin{abstract}
We systematically calculate thermopower of biased and unbiased
multilayer graphene systems. The effect of screening to a bias
field perpendicular to the graphene planes is taken into
account self-consistently under the Hartree approximation. The
model including nearest neighbor hopping and the more general
Slonczewski-Weiss-McClure (SWMcC) model are both considered for
a comparison. The effect of impurity scattering is studied for
monolayer and unbiased bilayer graphene and is treated in terms
of the self-consistent Born approximation. For a monolayer
graphene, only when the effect of impurity scattering is taken
into account, could all the qualitative aspects of the
experimental results be correctly reproduced. Besides bilayer
graphene, only trilayer graphene opens a small gap and shows a
slight enhancement of thermopower under an external bias. The
biased bilayer graphene shows the largest thermopower among all
the systems studied.
\end{abstract}


\maketitle

\section{\label{sec:Introduction}Introduction}

Recently, thermopower of graphene systems have attracted much
attention.\cite{zuev09,wei09,checkelsky09,lofwander07,hwang09,yan09,
ouyang09,dora07,zhu10,xing09,hao10,nam10,leeweili}
In experiments on monolayer graphene (MLG), while the high
carrier density region is well accounted for in terms of the
Mott's relationship, the low carrier density region shows
deviation away from this
behavior.\cite{zuev09,wei09,checkelsky09} This is explained in
terms of either an electron-hole puddle model\cite{hwang09} or
the coherence effect between the conduction band and the
valence band mediated by impurity scattering\cite{yan09}.
Besides the MLG, a bilayer graphene (BLG) system is also very
interesting. The opening of a gap by an external electric field
applied perpendicular to the layer planes
\cite{mccann06,min07,castro07,oostinga08,zhang09,mak09}
introduces a new degree of tunability to the system and is
recently shown by the authors to enhance greatly the
thermopower.\cite{hao10}

Despite the above progresses, several questions remain to be
answered. First, the dependence of thermopower on carrier
density as a function of temperature for a MLG has two features
which are not fully explained in previous works. The first one
is the shift of thermopower peak with respect to the charge
neutral point as temperature decreases. In addition,
thermopower shows a monotonic dependence on temperature for
carrier densities smaller than that of the thermopower
peak.\cite{ouyang09,hwang09,yan09} The second question is the
systematic dependence of thermopower on the number of layers.
Further more we are interested in finding out whether or not an
applied biased potential will enhance thermopower for the
multilayer graphene as the same as the bilayer system. Below we
shall address these questions.

Full microscopic calculations of thermopower of MLG in the
presence of charged impurity scattering are performed in terms
of the self-consistent Born approximation. Better agreement
with experiments\cite{zuev09,wei09,checkelsky09} as compared
with previously reported results are
achieved.\cite{hwang09,yan09,ouyang09} We also study
thermopower of impure unbiased bilayer graphene. Results which
are qualitatively similar to those in monolayer graphene are
obtained, but for much higher impurity concentrations.

For a multilayer graphene with layer number from two to six, we
consider both the nearest neighbor tight binding model and the
more general Slonczewski-Weiss-McClure (SWMcC)
model.\cite{slonczewski58,mcclure57} For a multilayer system
the effect of screening is important in determining the band
structure, as it produces non-uniform charge densities across
the
layers.\cite{mccann06,min07,koshino09a,avetisyan09,guinea07,ohta07,miyazaki08,lee09,sui09}
In this work, by using the self-consistent Hartree
approximation we obtain potential energies of different layers
for unbiased systems with three or more layers and biased
systems with a perpendicular electric field for two or more
layers. For unbiased graphene multilayers, peak values of the
thermopower are close to each other irrespective of the model
used. For a biased multilayer graphene, only thermopower of the
bilayer shows significant enhancement, thermopower of the
trilayers also increases slightly, but thermopower of systems
with more layers decreases under a bias. The huge enhancement
of thermopower in a bilayer graphene is a direct result of gap
opening under a perpendicular electric bias.\cite{hao10} In a
trilayer graphene, a much smaller gap opens and hence the
thermopower enhancement is very small. While for layer number
larger than three, no gap opens under a bias. Thus no
thermopower enhancement is expected for a multilayer graphene
with layer number larger than three.

\section{Model and Method}

\subsection{The SWMcC model}
We consider multilayer graphene systems stacked in the standard
AB (Bernal) stacking between consecutive
layers.\cite{nilsson08,partoens06,partoens07,hao09} In order to
correctly reproduce the band structure, a tight binding type of
model up to the fifth nearest neighbor hopping is
employed.\cite{slonczewski58,mcclure57,nilsson08,partoens06,partoens07,avetisyan09}
For an $N$-layer graphene, the Hamiltonian is transformed in
the $xy$-plane to wave vector space, whereas the direction
perpendicular to the plane is kept in the real space. Along the
$z$-direction, the lattice sites which are vertically aligned
to each other are labeled as ($A_1$, $B_2$, $A_3$, ...), with
$A_i$ ($B_i$) denoting the $A$ ($B$) sublattice atoms on the
$i$-th layer.\cite{hao09} The single spin Hamiltonian is thus
written as\cite{nilsson08,partoens06,partoens07,avetisyan09}
\begin{eqnarray}
&&H_{N}=\sum\limits^{N}_{n=1}\sum\limits_{\mathbf{k}}
[\phi(\mathbf{k})a^{\dagger}_{n\mathbf{k}}b_{n\mathbf{k}}+H.c.] \notag \\
&&+\sum\limits_{n,\mathbf{k}}[(\Delta+\gamma_5)(a^{\dagger}_{2n-1,\mathbf{k}}a_{2n-1,\mathbf{k}}
+b^{\dagger}_{2n,\mathbf{k}}b_{2n,\mathbf{k}}) \notag \\
&&+\gamma_{2}(b^{\dagger}_{2n-1,\mathbf{k}}b_{2n-1,\mathbf{k}}+a^{\dagger}_{2n,\mathbf{k}}a_{2n,\mathbf{k}})]  \notag \\
&& +\sum\limits_{n,\mathbf{k}}[\gamma_{1}(a^{\dagger}_{2n-1,\mathbf{k}}b_{2n,\mathbf{k}}
+a^{\dagger}_{2n+1,\mathbf{k}}b_{2n,\mathbf{k}}+H.c.)  \notag \\
&&+\frac{\gamma_{5}}{2}(a^{\dagger}_{2n-1,\mathbf{k}}a_{2n+1,\mathbf{k}}+b^{\dagger}_{2n,\mathbf{k}}b_{2n+2,\mathbf{k}}+H.c.)  \notag \\
&&+\frac{\gamma_{2}}{2}(b^{\dagger}_{2n-1,\mathbf{k}}b_{2n+1,\mathbf{k}}+a^{\dagger}_{2n,\mathbf{k}}a_{2n+2,\mathbf{k}}+H.c.)] \notag \\
&&+v_{3}\sum\limits_{n,\mathbf{k}}[\phi(\mathbf{k})b^{\dagger}_{2n-1,\mathbf{k}}a_{2n,\mathbf{k}}
+\phi(\mathbf{k})b^{\dagger}_{2n+1,\mathbf{k}}a_{2n,\mathbf{k}}+H.c.]  \notag \\
&&-v_{4}\sum\limits_{n,\mathbf{k}}[\phi^{\ast}(\mathbf{k})a^{\dagger}_{2n-1,\mathbf{k}}a_{2n,\mathbf{k}}
+\phi^{\ast}(\mathbf{k})b^{\dagger}_{2n-1,\mathbf{k}}b_{2n,\mathbf{k}}  \notag \\
&&+\phi^{\ast}(\mathbf{k})a^{\dagger}_{2n+1,\mathbf{k}}a_{2n,\mathbf{k}}+\phi^{\ast}(\mathbf{k})b^{\dagger}_{2n+1,\mathbf{k}}b_{2n,\mathbf{k}}+H.c.].
\end{eqnarray}
In the above expression,
$\phi(\mathbf{k})$=$\gamma_{0}\sum\limits^{3}_{l=1}\exp(i\mathbf{k}\cdot\boldsymbol{\delta}_{l})$
arises from motion within separate layers.
$v_{3}$=$\gamma_{3}/\gamma_{0}$,
$v_{4}$=$\gamma_{4}/\gamma_{0}$. $\gamma_{0}$, $\gamma_{1}$,
$\gamma_{2}$, $\gamma_{3}$, $\gamma_{4}$, $\gamma_{5}$ and
$\Delta$ are standard SWMcC parameters, and are taken as 3.12
eV, 0.377 eV, -0.02 eV, 0.29 eV, 0.12 eV, 0.0125 eV and 0.004
eV,
respectively.\cite{slonczewski58,mcclure57,partoens06,partoens07,nilsson08}
Except the first term, summation over $n$ runs from 1 to $N/2$
or $N/2 \pm 1$ etc., so that  the basis vector
$\psi^{\dagger}(\mathbf{k})$=($a^{\dagger}_{1\mathbf{k}}$,
$b^{\dagger}_{1\mathbf{k}}$; $a^{\dagger}_{2\mathbf{k}}$,
$b^{\dagger}_{2\mathbf{k}}$;...; $a^{\dagger}_{N\mathbf{k}}$,
$b^{\dagger}_{N\mathbf{k}}$) has $2N$ components. The
Hamiltonian is simply given as
\begin{equation}
H_{N}=\sum\limits_{\mathbf{k}}\psi^{\dagger}_{\mathbf{k}}H(\mathbf{k})\psi_{\mathbf{k}},
\end{equation}
where matrix form of $H(\mathbf{k})$ is not shown explicitly
here but can be found in Refs. \cite{avetisyan09} and
\cite{partoens07}.

\subsection{Nearest neighbor tight binding model}
Apart from the full SWMcC model, a simplified model up to
nearest neighbor intralayer and interlayer hopping is widely
used to study many problems. In this case, the Hamiltonian
matrix is written
as\cite{koshino06,nilsson08,guinea06,zhang09b,hao09}
\begin{equation} \label{h2}
H(\mathbf{k})=\begin{pmatrix} H_{0}(\mathbf{k}) & V &  &  &  \\ V^{\dagger} & H_{0}(\mathbf{k}) &
V^{\dagger} &  &  \\  & V & H_{0}(\mathbf{k}) & V &  \\  &  & \ddots & \ddots &
\ddots
\end{pmatrix},
\end{equation}
in which
\begin{equation}
H_{0}(\mathbf{k})=\begin{pmatrix} 0 & \phi(\mathbf{k}) \\ \phi^{*}(\mathbf{k}) & 0
\end{pmatrix}, V=\begin{pmatrix} 0 & \gamma_{1} \\ 0 & 0 \end{pmatrix}.
\end{equation}
In the absence of any external field or gate voltage, it is
known that the above Hamiltonian could be decomposed into
subsystems of bilayer and monolayer
graphene.\cite{koshino06,koshino07,koshino08,hao09} For an
odd-layered multilayer graphene, there are one MLG subsystem
and $(N-1)/2$ different BLG subsystems. While for an
even-layered multilayer, there are $N/2$ different bilayers.
Label the different BLG subsystems with an index $m$, and take
the basis vector as
$\psi^{\dagger}_{m}(\mathbf{k})$=($a^{\dagger}_{m1\mathbf{k}}$,
$b^{\dagger}_{m1\mathbf{k}}$, $a^{\dagger}_{m2\mathbf{k}}$,
$b^{\dagger}_{m2\mathbf{k}}$), the Hamiltonian matrix of the
$m$-th BLG subsystem is written as
\begin{equation} \label{subblg}
H_m(\mathbf{k})=\begin{pmatrix} 0 & \phi(\mathbf{k}) & 0 & \gamma_{1m} \\
\phi^{\ast}(\mathbf{k}) & 0 & 0 & 0 \\ 0 & 0 & 0 &
\phi(\mathbf{k}) \\ \gamma_{1m} & 0 & \phi^{\ast}(\mathbf{k}) & 0 \end{pmatrix},
\end{equation}
where $\gamma_{1m}$=$2\gamma_{1}\sin{\frac{m\pi}{2(N+1)}}$. $m$
takes the value of $N-1$, $N-3$, $\cdots$, 2 (for odd $N$) or 1
(for even $N$).

\subsection{Screening effects}
In our former work, a bias voltage is shown to greatly enhance
the thermopower of BLG.\cite{hao10} It is unclear whether this
is true also for other multilayer graphene. In the presence of
a bias voltage or a charged gate, the potential energy
difference would be induced between different layers. To
determine the distribution of potential energies,
self-consistent calculations are required to take into account
of the Coulomb screening arising from charge redistribution
among the different
layers.\cite{mccann06,guinea07,avetisyan09,koshino10} For
unbiased multilayer graphene with layer number larger than two,
the charge and potential energy distribution should also be
obtained self-consistently.

There are two schemes to account for the
charge redistribution. One is to consider the screening of a
charged gate with fixed carrier density by the multilayer
graphene system\cite{avetisyan09}. The other approach is to
consider the charge redistribution problem of the multilayer
graphene system in the presence of a perpendicular electric
field, while the carrier density is controlled by another gate
voltage applied equally to all layers\cite{koshino10}. Here we
follow the second approach.

Suppose an external electric field $\mathbf{E}_0$ is applied to
the free-standing multilayer graphene system along the direction
perpendicular to the graphene planes from layer 1 to layer $N$
($\mathbf{E}_0$=0 for unbiased systems).
After establishing equilibrium, the $i$-th layer has a
total excess electron density of $n_{i}$. According to Gauss's
law, the screening electric field pointing from the $j$-th to
the $(j$+$1)$-th layer is
$E^{sc}_{(j,j+1)}$=$\frac{e}{2\epsilon_{r}}(\sum\limits^{N}_{i=j+1}n_{i}-\sum\limits^{j}_{i'=1}n_{i'})$,
where `$sc$' means screening, $e$ is the absolute value of the electron charge,
and $\epsilon_{r}$=3 is the static dielectric constant
of the multilayer graphene\cite{yan09}.
The total electric field is thus $E_{(j,j+1)}$=$E_{0}+E^{sc}_{(j,j+1)}$.
Introduce the doping $x_{i}$=$n_{i}\Omega_{0}$
($\Omega_{0}$=$\frac{3\sqrt{3}}{4}a^{2}$ is the effective area
per carbon site), which means the number of excess electrons
per site residing on the $i$-th layer. Define two parameters
$V_{0}$=$eE_{0}d_{0}$ and
$\gamma$=$\frac{e^{2}d_{0}}{2\epsilon_{r}\Omega_{0}}$.
$d_{0}$$\simeq$3.5 \AA\ is the vertical distance between adjacent layers.
The potential energy difference between the $j$-th and the
$(j$+$1)$-th layer is
\begin{equation}  \label{dV}
\Delta V_{j}=V_{j+1}-V_{j}=V_{0}+\gamma[\sum\limits^{N}_{i=j+1}x_{i}-\sum\limits^{j}_{i'=1}x_{i'}].
\end{equation}
The total difference of on-site potential energy is thus
\begin{eqnarray}
&&V_{N}-V_{1}=\sum\limits^{N-1}_{j=1}\Delta V_{j} \notag \\
&&=(N-1)V_{0}+\gamma\sum\limits^{N-1}_{j=1}(N-j)(x_{N-j+1}-x_{j}).
\end{eqnarray}
Since only differences in the potential energies are relevant,
we would take the potential energy of the first and $N$-th
layer symmetric with respect to zero by shifting the energy
reference. Thus we take $\tilde{V}_{N}$=$(V_{N}-V_{1})/2$ and
$\tilde{V}_1$=-$\tilde{V}_N$. Potential energy differences
$\tilde{V}_{j+1}-\tilde{V}_{j}$ between consecutive layers are
not influenced by this shift of energy reference, and are still
determined by Eq. (\ref{dV}).

For a certain $V_{0}$ and a fixed average doping, we calculate
the set of doping \{$x_{i}$\} self-consistently. Label the
2$N$$\times$2$N$ matrix which diagonalizes the Hamiltonian
matrix as $U(\mathbf{k})$, that is
$U^{\dagger}(\mathbf{k})H(\mathbf{k})U(\mathbf{k})$=$H_{d}(\mathbf{k})$.
Here, we have added the on-site potential energies
$H_{V}$=$\sum\limits_{\mathbf{k}}\psi^{\dagger}_{\mathbf{k}}H_{V}(\mathbf{k})\psi_{\mathbf{k}}$
to the model and have kept the notation of the Hamiltonian
matrix unchanged as $H(\mathbf{k})$. The diagonal matrix
$H_{V}(\mathbf{k})$ is $\mathbf{k}$ independent and is defined
as diag[$\tilde{V}_{1}$, $\tilde{V}_{1}$, ... ,
$\tilde{V}_{N}$, $\tilde{V}_{N}$]. The $\alpha$-th column of
$U(\mathbf{k})$ stores the $\alpha$-th eigenvector of
$H(\mathbf{k})$ corresponding to an eigenenergy of
$\epsilon_{\alpha}(\mathbf{k})$=$[H_{d}(\mathbf{k})]_{\alpha\alpha}$.
Suppose the number of wave vectors considered in the Brillouin
zone is $N_{\mathbf{k}}$, the self-consistency condition for
determining $x_{i}$ is thus
\begin{equation}
x_{i}+1=\frac{1}{N_{\mathbf{k}}}\sum\limits_{\mathbf{k},\alpha}(|U_{2i-1,\alpha}(\mathbf{k})|^2+|U_{2i,\alpha}(\mathbf{k})|^2)
f(\epsilon_{\alpha}(\mathbf{k})-\mu).
\end{equation}
$\mu$ is the chemical potential. $f(x)$=$1/(e^{\beta x}+1)$ is
the Fermi distribution function, in which $\beta$ represents
the inverse temperature $1/k_{B}T$ with $k_{B}$ denoting the
Boltzmann constant. After a set of convergent results for
$\{x_i, \tilde{V}_i; i=1, ..., N\}$ is obtained, the
corresponding on site potential energies $\tilde{V}_{i}$ ($i=1,
..., N$) are substituted into the model to calculate the
thermopower of interest in this work.

\subsection{Scattering by charged impurities}
Transport properties of graphene systems sensitively depend on
the nature and concentration of impurities. In particular,
charged impurities are shown to dominate the various transport
properties of MLG\cite{adam07,hwang07,hwang07b,nomura07} and
also play a very important role in the transport of
BLG\cite{xiao09}. In the following, we would consider the
effect of charged impurity scattering in the MLG and unbiased
BLG systems. The treatment would be at the level of
self-consistent Born approximation
(SCBA).\cite{lee93,shon98,koshino06,yan09,yan08,peres06,ostrovsky06}
Due to the limitation in computation time and the fact that
calculations performed for clean systems are enough to answer
if a bias could enhance thermopower of a multilayer graphene,
in this work we would concentrate on clean systems for layer
number larger than two.

The formalism is presented in detail in our former work on
thermopower of gapped BLG\cite{hao10}. The thermopower is
represented in terms of the linear response coefficients
as\cite{hwang09,yan09,jonson80,hao10}
\begin{equation} \label{S}
S=-\frac{L_{12}}{eTL_{11}},
\end{equation}
where $T$ is the absolute temperature and $e$ is the magnitude
of electron charge. The linear response coefficients are
obtained as
\begin{equation}
\mathit{L}_{ij}=\lim_{\omega\rightarrow
0}\text{Re}\mathcal{L}_{ij}(\omega+i0^{+}).
\end{equation}
In the Matsubara notation, the correlation function
reads\cite{jonson80}
\begin{equation}
\mathcal{L}_{ij}(i\omega_{n})=-\frac{iT}{(i\omega_{n})\Omega d}
\int_{0}^{\beta} d\tau e^{i\omega_{n}\tau}\langle
T_{\tau}\mathbf{j}_{i}(\tau)\cdot\mathbf{j}_{j}(0)\rangle,
\end{equation}
The two subindices `i' and `j' both run over 1 and 2. $\Omega$
is the total effective area of the system defined as
the area per layer multiplied by $N$, the number of layers.
$d$=2 is the dimensionality.
$\mathbf{j}_{1}$ and $\mathbf{j}_{2}$ denote particle current
and heat current operators, respectively.\cite{hao10} The two
linear response coefficients in terms of which the thermopower
is obtained are written as
\begin{eqnarray} \label{L1j}
L_{1j}=& &T\int_{-\infty}^{+\infty}\frac{d\epsilon}{2\pi}
[-\frac{\partial f(\epsilon)}{\partial \epsilon}]\text{Re}\{P_{1j}(\epsilon-i0^{+},\epsilon+i0^{+}) \notag\ \\
& & -P_{1j}(\epsilon+i0^{+},\epsilon+i0^{+})\},
\end{eqnarray}
where the subindex $j$ is either $1$ or $2$. The kernels
are defined as\cite{hao10}
\begin{equation}
P_{1j}(z,z')=\epsilon^{j-1}\frac{2}{\Omega d}\sum\limits_{\mathbf{k}}\text{Tr}
\{G_{\mathbf{k}}(z)\boldsymbol{\Gamma}_{1}(\mathbf{k},z,z')G_{\mathbf{k}}(z')\cdot\mathbf{j}_{1}^{\mathbf{k}}\}.
\end{equation}
with $\boldsymbol{\Gamma}_{1}(\mathbf{k},z,z')$ as the vertex
function corresponding to the wave vector $\mathbf{k}$. Here,
$z$=$\epsilon\pm i0^{+}$ and $z'$=$\epsilon+i0^{+}$.
$\mathbf{j}^{\mathbf{k}}_{1}$ is the matrix for the charge
current at wave vector $\mathbf{k}$ and is obtained from the
Hamiltonian matrix $H(\mathbf{k})$ of the multilayer graphene
as\cite{ambegaokar65,durst00,hao10}
\begin{equation}
\mathbf{j}^{\mathbf{k}}_{1}=\boldsymbol{\nabla}_{\mathbf{k}}H(\mathbf{k}).
\end{equation}
In $H(\mathbf{k})$ the potential energy distribution
$H_{V}(\mathbf{k})$ with the on site energies determined
self-consistently are incorporated, which affects the energy
spectrum but does not affect the current operator.

For an $N$-layer multilayer graphene, the Green's function is
represented in terms of a $2N$$\times$$2N$ matrix as
\begin{equation}
G_{\mathbf{k}}(z)=[G^{0}_{\mathbf{k}}(z)^{-1}-\Sigma_{\mathbf{k}}(z)]^{-1}.
\end{equation}
The free Green's function is obtained as
$G_{\mathbf{k}}^{0}(z)=[(z+\mu)I_{2N}-H(\mathbf{k})]^{-1}$,
where $I_{2N}$ is a $2N$-dimensional unit matrix and $\mu$ is
the chemical potential determined from $H(\mathbf{k})$ for a
certain carrier density and temperature. The self energies and
the vertex functions are determined self-consistently in terms
of the following iteration functions\cite{yan09,hao10}
\begin{equation}
\Sigma_{\mathbf{k}}(z)=\frac{n_{i}}{\Omega}\sum\limits_{\mathbf{k}'}|v_{i}(\mathbf{k-k'})|^{2}
[G_{\mathbf{k}'}^{0}(z)^{-1} -\Sigma_{\mathbf{k}'}(z)]^{-1},
\end{equation}
\begin{eqnarray}
&&\boldsymbol{\Gamma}_{1}(\mathbf{k},z,z')=\mathbf{j}_{1}^{\mathbf{k}} \\
&&+\frac{n_{i}}{\Omega}
\sum\limits_{\mathbf{k'}}|v_{i}(\mathbf{k-k'})|^{2}G_{\mathbf{k}'}(z)\boldsymbol{\Gamma}_{1}(\mathbf{k'},z,z') G_{\mathbf{k}'}(z').  \notag
\end{eqnarray}
$n_{i}$ is the impurity concentration averaged to per layer.
$v_{i}(\mathbf{q})$ is the electron-impurity scattering
potential. For charged impurity scattering, $v_{i}(\mathbf{q})$
is taken as of the Thomas-Fermi type\cite{yan08,hwang09,yan09}
\begin{equation}
v_{i}(\mathbf{q})=\frac{2\pi e^{2}}{\epsilon_{r}(q+q_{TF})}e^{-qd_{i}}.
\end{equation}
$\epsilon_{r}$ is the effective dielectric constant from
lattice and substrate, $\epsilon_{r}$=3 is adopted in this
work\cite{hwang07b,yan08,yan09}. $d_{i}$ is the distance between
the impurities and the graphene plane and would be set as zero
in the present work\cite{hwang09,yan09}. $q_{TF}$ is the
Thomas-Fermi wave number and is obtained from the
long-wavelength-limit static polarizability of the
corresponding noninteracting electron
system\cite{yan08,hwang09} as
\begin{equation}
q_{TF}=2\pi e^2\chi/\epsilon_{r},
\end{equation}
with the static polarizability
\begin{equation}
\chi=\frac{2}{\Omega}\int_{0}^{\beta}d\tau\langle\text{T}_{\tau}n(\tau)n^{\dagger}(0)\rangle_{c}.
\end{equation}
The particle number operator is defined as
$n(\tau)$=$\sum\limits_{\mathbf{k}}\psi_{\mathbf{k}}^{\dagger}(\tau)\psi_{\mathbf{k}}(\tau)$.
A factor of `2' comes from the two fold degeneracy in spin. The
subindex `$c$' means retaining only connected Feynman diagrams
in evaluating the expectation value.\cite{hao10}

The self-consistent Born approximation (SCBA) enters when
making averages over impurity configurations as\cite{mahanbook}
\begin{equation}
\langle\rho_{i}(\mathbf{q})\rho_{i}(-\mathbf{q}')\rangle=N_{i}\delta_{\mathbf{q,q'}},
\end{equation}
where $N_{i}=n_{i}\Omega$ is number of impurities in the system
under consideration.

For clean systems, $G_{\mathbf{k}}(z)$ reduces to the free
Green's function and the vertex function
$\boldsymbol{\Gamma}_{1}(\mathbf{k},z,z')$ reduces to the bare charge
current matrix $\mathbf{j}_{1}^{\mathbf{k}}$.\cite{hao10}

\section{result and discussion}

\subsection{Monolayer graphene}
The experimental results\cite{zuev09,wei09,checkelsky09} for
thermopower of a MLG show two interesting features. First, for
a certain temperature, thermopower follows $1/\sqrt{|x|}$ ($x$
is the average number of excess electrons per site) for high
carrier densities but then the magnitude decreases and changes
sign close to the charge neutrality point. Theoretical
calculations in terms of both the Boltzmann transport
equation\cite{hwang09} and the microscopic Kubo's
formula\cite{yan09} have successfully reproduced the above
features. The deviation of thermopower from the $1/\sqrt{|x|}$
behavior close to the charge neutrality point is ascribed to
electron-puddle formation\cite{hwang09} or coherence between
the conduction band and valence band\cite{yan09}, which both
imply the coexistence of electron-like and hole-like characters
of the carriers. The second feature is that, as temperature
decreases the thermopower decreases monotonously for all
carrier densities and the peak position also shifts toward
lower carrier densities.\cite{zuev09,wei09} The above behaviors
are very similar to the results presented in Figs. 1(c) and
1(d).

Results in Fig. 1 are obtained in terms of the nearest neighbor
hopping model with hopping integral as 2.7 eV. As shown in Fig.
1(a) and 1(b), calculations on clean systems are unable to
reproduce the second feature mentioned above.\cite{ouyang09}
Previous works on transport properties of a MLG had confirmed
the importance of charged
impurities.\cite{hwang07,adam07,hwang07b,nomura07} Here, we
treat the impurity scattering in terms of
SCBA.\cite{shon98,yan09} We would neglect the shift of chemical
potential by the impurity potential.\cite{yan09,hao10} The
results for a series of temperatures with impurity
concentration $n_{i}=10^{12}$ cm$^{-2}$, which corresponds to a
moderately disordered sample\cite{adam07}, are shown in Fig.
1(c). For the electron-hole symmetric band as considered here,
the relationship $S(-x)=-S(x)$ generally holds. The full curve
for 300 K in Fig. 1(c) is explicitly calculated, the above
relationship is seen to hold very well. For all other curves,
only the hole doping part with $x\le 0$ is calculated
explicitly and the part with $x>0$ is obtained in terms of
$S(x)=-S(-x)$. It is clear that, Fig. 1(c) agrees qualitatively
very well with experiment by Zuev \emph{et al.}\cite{zuev09} As
shown in Fig. 1(d), increasing $n_{i}$ to $5\times10^{12}$
cm$^{-2}$, which corresponds to dirty graphene samples, the
peak positions are shifted to higher carrier densities and the
maximum values of thermopower are further suppressed for all
temperatures, which are qualitatively similar to the
experimental results by Wei \emph{et al.}\cite{wei09}

\begin{figure}
\centering
\includegraphics[width=9cm,height=8cm,angle=0]{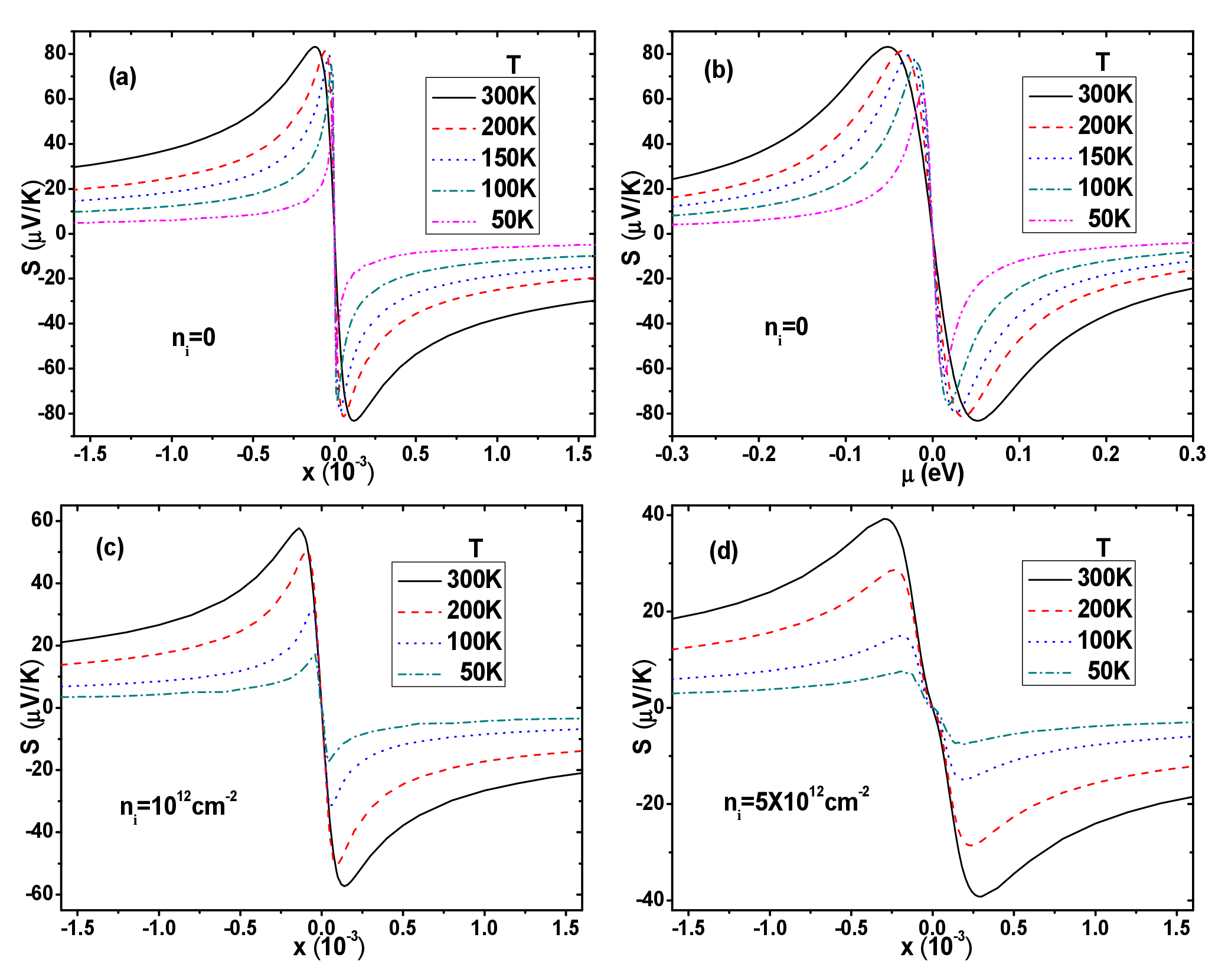}
\caption{Thermopower of clean monolayer graphene at five temperatures as a function of
(a) carrier density and (b) chemical potential. Thermopower of impure monolayer graphene
at four temperatures for two different impurity
concentrations as (c) 10$^{12}$ cm$^{-2}$ and (d) 5$\times$10$^{12}$ cm$^{-2}$.}
\end{figure}

In the degenerate region for relatively high doping
concentrations, the semiclassical Mott's relationship generally
holds.\cite{cutler69,hwang09} In these cases, for a certain
carrier density (or, gate voltage), the low temperature
thermopower scales linearly with temperature. Experimentally,
the Mott's relationship holds very well for graphene at high
carrier densities but breaks down close to the charge
neutrality point. Fig. 2 shows our results for three impurity
concentrations. The linear temperature dependence holds even
for doping close to (but still larger than) the peak position.
For carrier densities smaller than at the peak position (e.g.,
$x$$\simeq$$\pm 1.4\times 10^{-4}$ for T=300 K and
$n_{i}$=10$^{12}$ cm$^{-2}$) where electron-like and hole-like
carriers coexist, the linear Mott's relationship is not valid.
As impurity concentration increases, the peak position shifts
to higher carrier densities, region of carrier density for the
violation of linear temperature dependence is correspondingly
enlarged.

\begin{figure}
\centering
\includegraphics[width=7cm,height=11cm,angle=0]{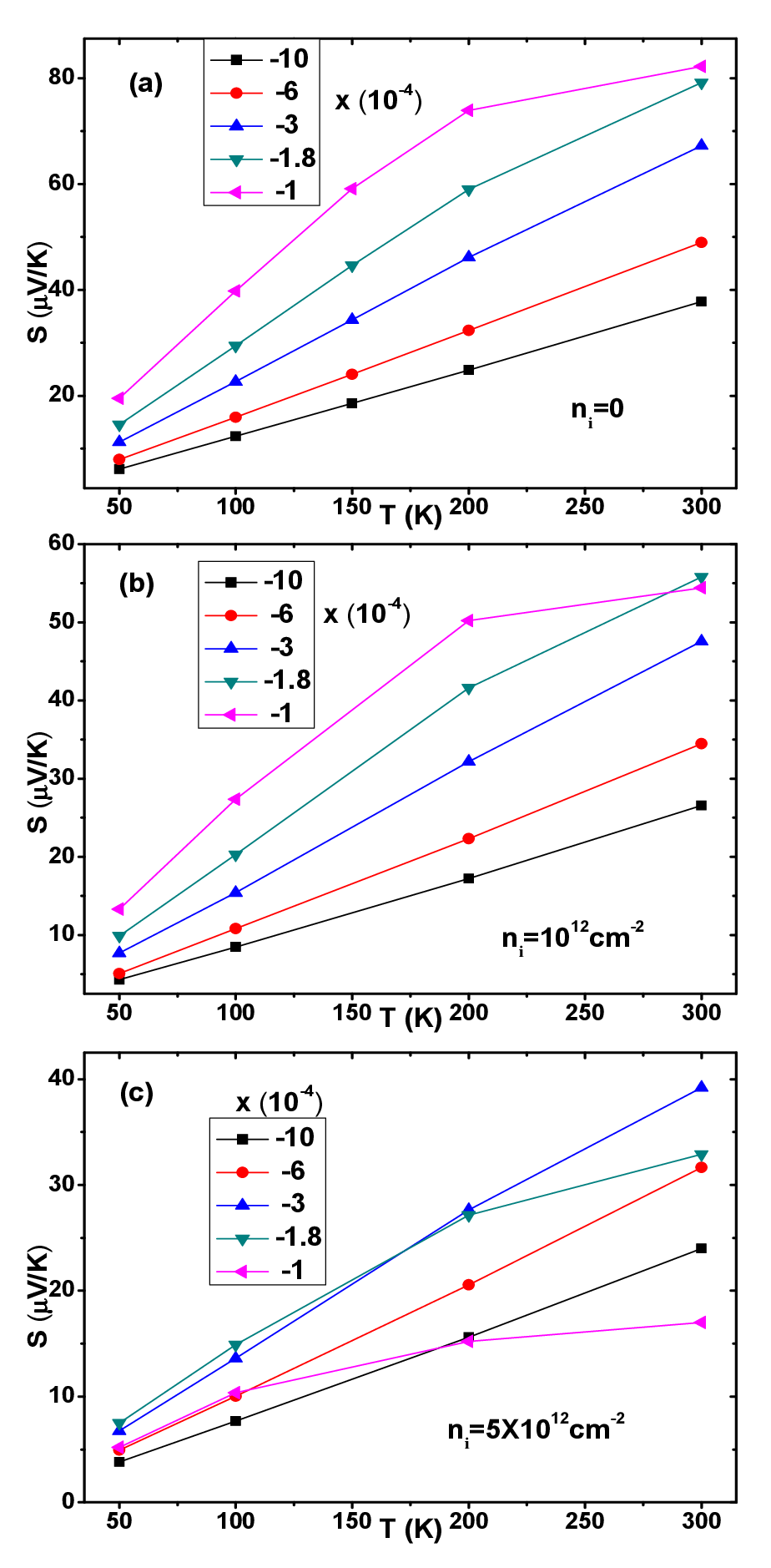}
\caption{Thermopower of monolayer graphene as a function of temperature is plotted for five electron densities
and for  (a) the clean system.
(b) $n_{i}$=10$^{12}$ cm$^{-2}$. (c) $n_{i}$=5$\times$10$^{12}$ cm$^{-2}$.}
\end{figure}

We would like to compare our results with previous calculations
taking into account of scattering by charged impurities. Hwang
\emph{et al.}\cite{hwang09} study the problem in terms of
semiclassical Boltzmann equation approach. There, the high and
low carrier density regions are treated separately. Hence it is
difficult to provide the dependence of peak position with
temperature. Another work by Yan \emph{et al.}\cite{yan09}
starts from the microscopic Kubo's formula but has only focused
on the low temperature limit and is also unable to reproduce
the temperature dependence of the peak position observed
experimentally\cite{zuev09,wei09,checkelsky09}. Compared to the
above two works, we have shown that in terms of fully
microscopic calculations incorporating the effect of scattering
by charged impurities, both the two features mentioned in the
beginning of this section could be reproduced successfully. Our
results also confirm the picture that the peak position shifts
to higher carrier density with the increase of both temperature
and impurity concentration.\cite{yan09,hao10}.

\subsection{Unbiased bilayer graphene: The influence of impurity scattering}
Having achieved a reasonable success by using the SCBA to treat
scatterings by charged impurities in a MLG, we shall proceed to
study the bilayer system. We have previously studied the effect
of charged impurity scattering on the thermopower of gapped
(biased) BLG\cite{hao10}. In a gapped BLG, localization effect
becomes important in the low carrier density region, which is
out of the reach of
SCBA.\cite{nilsson07,koshino08b,mkhitaryan08} There we
restricted our attention to the systems with dilute impurity
concentrations.\cite{hao10} However, in an unbiased BLG with a
finite carrier density at the Fermi surface we expect to have a
much stronger screening which allows us to consider much larger
impurity concentrations.

\begin{figure}
\centering
\includegraphics[width=7cm,height=11cm,angle=0]{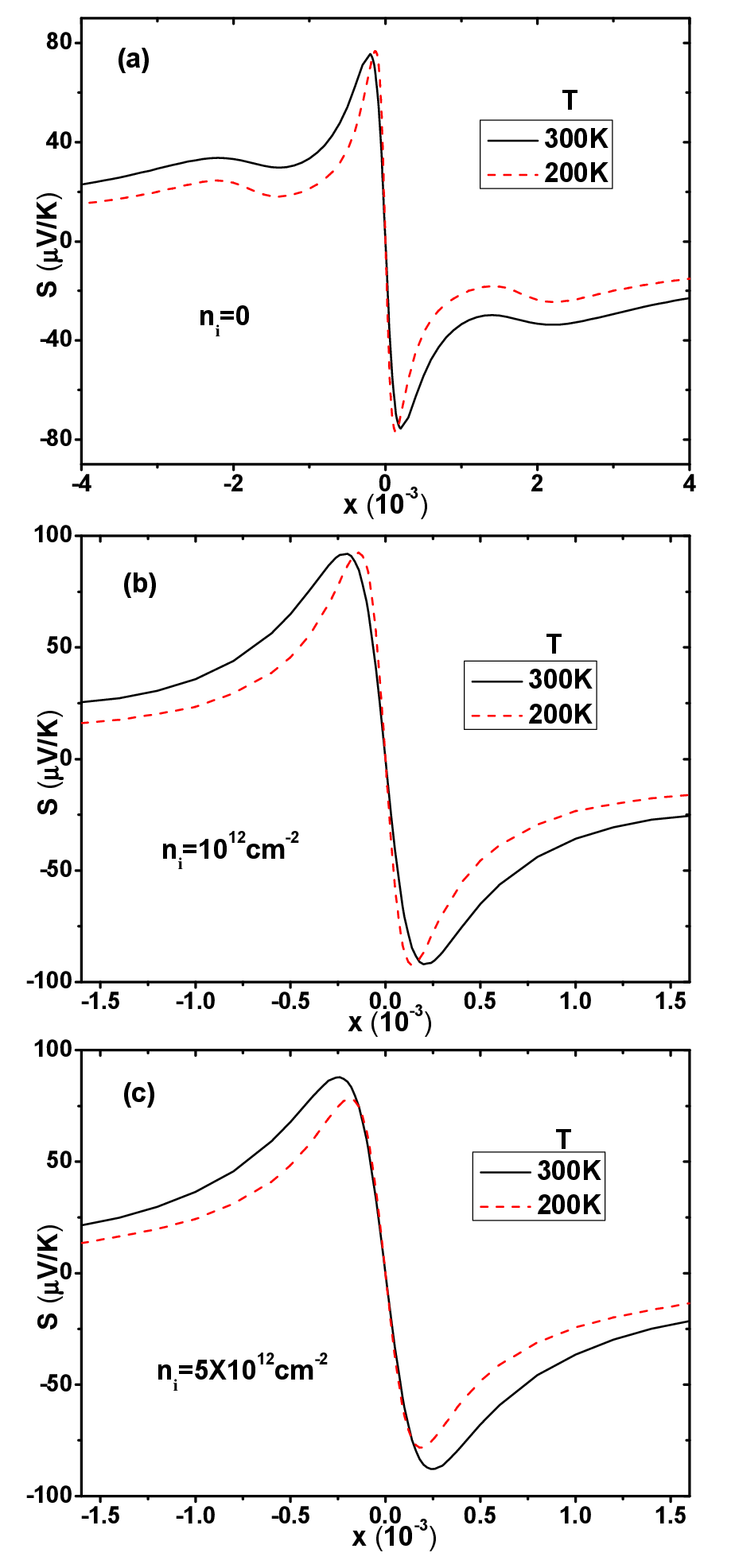}
\caption{Thermopower of unbiased bilayer graphene at two temperatures for (a) the clean system,
and the impure system at two different impurity
concentrations as (b) 10$^{12}$ cm$^{-2}$ and (c) 5$\times$10$^{12}$ cm$^{-2}$.}
\end{figure}

As a comparison to the results for MLG, we present in Fig. 3
thermopower of clean BLG and impure BLG for two typical
impurity concentrations: $10^{12}$ cm$^{-2}$ and
5$\times$10$^{12}$ cm$^{-2}$, obtained in terms of the nearest
neighbor hopping model with $\gamma_{0}$=3 eV and
$\gamma_{1}$=0.3 eV.\cite{hao10} For each impurity
concentration, the carrier density dependence of thermopower
for two typical temperatures, 300 K, 200 K are considered.
Lower temperature requires much larger number of wave vectors
to converge, which are difficult to approach. In contrast to
the MLG, impurity concentration up to $10^{12}$ cm$^{-2}$ does
not change the qualitative aspects of the results. Only when
the impurity concentration is increased to be as large as
5$\times$10$^{12}$ cm$^{-2}$, the result becomes similar to
that observed in experiments.\cite{nam10,leeweili} In both
experiments, the samples show semiconducting like transport
behaviors with a low mobility\cite{nam10,leeweili}, indicating
appreciable impurity concentrations and are thus in agreement
with our results.

The result of Fig. 3(c) is very similar to that of the Fig.
1(c) for a MLG except with an impurity concentration five times
larger. This is due to a much better screening in a bilayer
system. We present in Fig. 4 the Thomas-Fermi screening wave
vectors of both systems at two different temperatures: 300 K
and 50 K. At 300 K, $q_{TF}$ of BLG is more than five times
that of MLG around the charge neutral point. This difference
becomes even larger at 50 K. For the carrier density where the
thermopower peaks ($x\sim$$\pm$2$\times$10$^{-4}$) for BLG,
$q_{TF}$ of BLG is still more than two times that of MLG.
According to Eq. (16), Eq. (17), and Eq. (18), the effective
strength of impurity scattering is approximately proportional
to $n_{i}q_{TF}^{-2}$. Hence a BLG with a much larger
concentration of charged impurity shows the same qualitative
behavior as observed in MLG with a smaller impurity
concentration.

\begin{figure}
\centering
\includegraphics[width=8cm,height=7cm,angle=0]{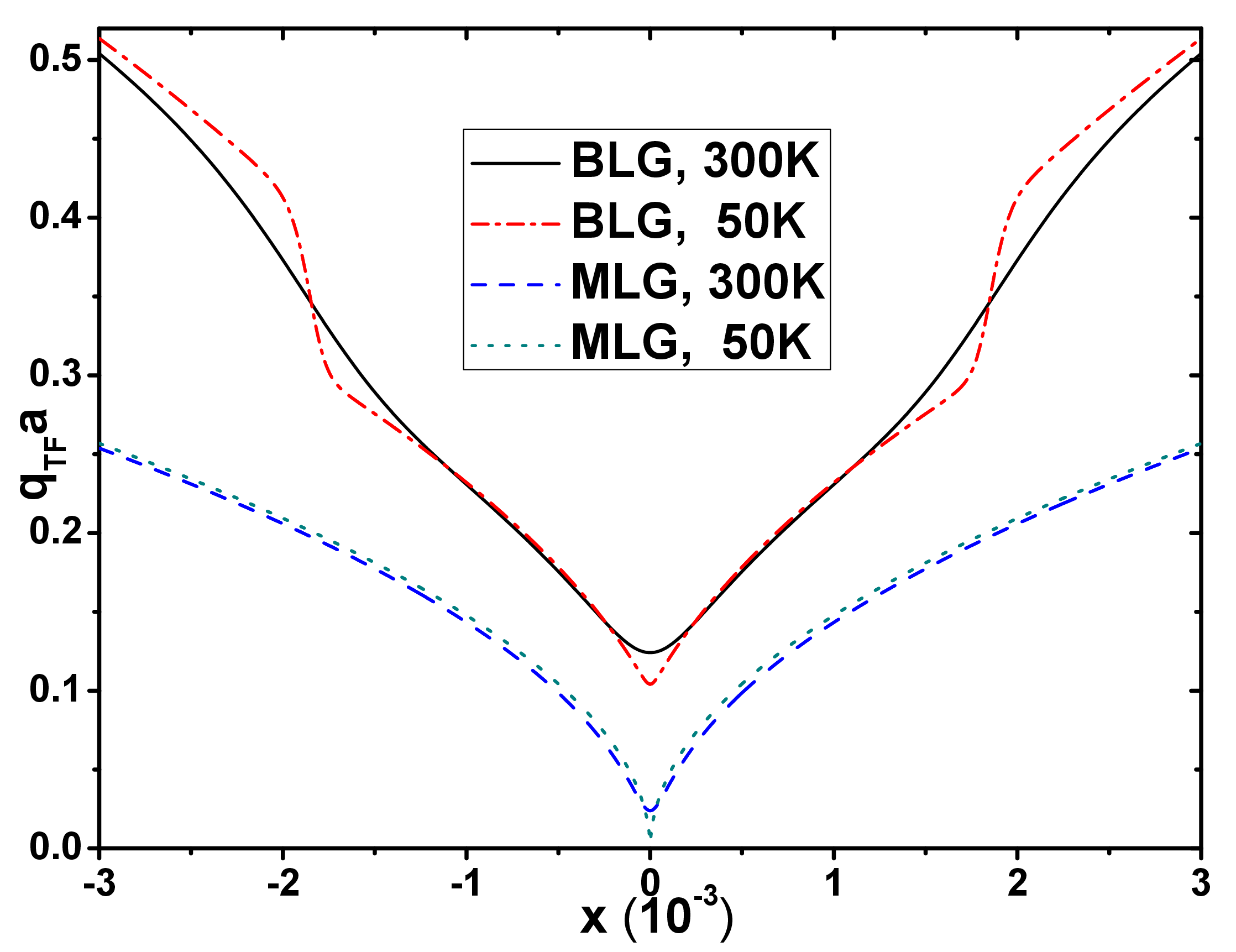}
\caption{Thomas-Fermi screening wave vector for an unbiased BLG
and a MLG at 300 K and 50 K.}
\end{figure}

The stronger screening in BLG as compared to that in MLG has
been used to make the conjecture that short-range scatterers
may also play an important role in the transport of gapless BLG
in addition to the charged impurities.\cite{sarma10} It would
be interesting to see how short-range scatterers would change
the results presented above.

\subsection{Unbiased multilayer graphene}
Now, we study the evolution of thermopower as a function of
layer number for an unbiased multilayer graphene. From now on
we would only consider clean graphene multilayers. First, we
consider the simple nearest-neighbor-hopping model, ignoring
temporarily the screening effect. As shown in previous works,
this simplified multilayer graphene model could be decomposed
into independent subsystems of MLG and
BLG.\cite{koshino06,koshino07,koshino08,hao09} The two linear
response coefficients $L_{11}$ and $L_{12}$ are then obtained
as summations of $L^{(m)}_{11}$ and $L^{(m)}_{12}$, which are
the corresponding values for the various subsystems labeled by
$m$.\cite{hao09} Thermopower of multilayers is then obtained
from Eq. (\ref{S}). The room temperature thermopowers obtained
in this way are shown in Fig. 5(a) with layer number up to six.
Peak value of the thermopower does not show very large
variation with layer number. Different from MLG, thermopower
for larger layer number samples usually show secondary peaks
associated with the onset of higher conduction or valence bands
contributing to transport.\cite{hao10}

\begin{figure}
\centering
\includegraphics[width=9cm,height=13cm,angle=0]{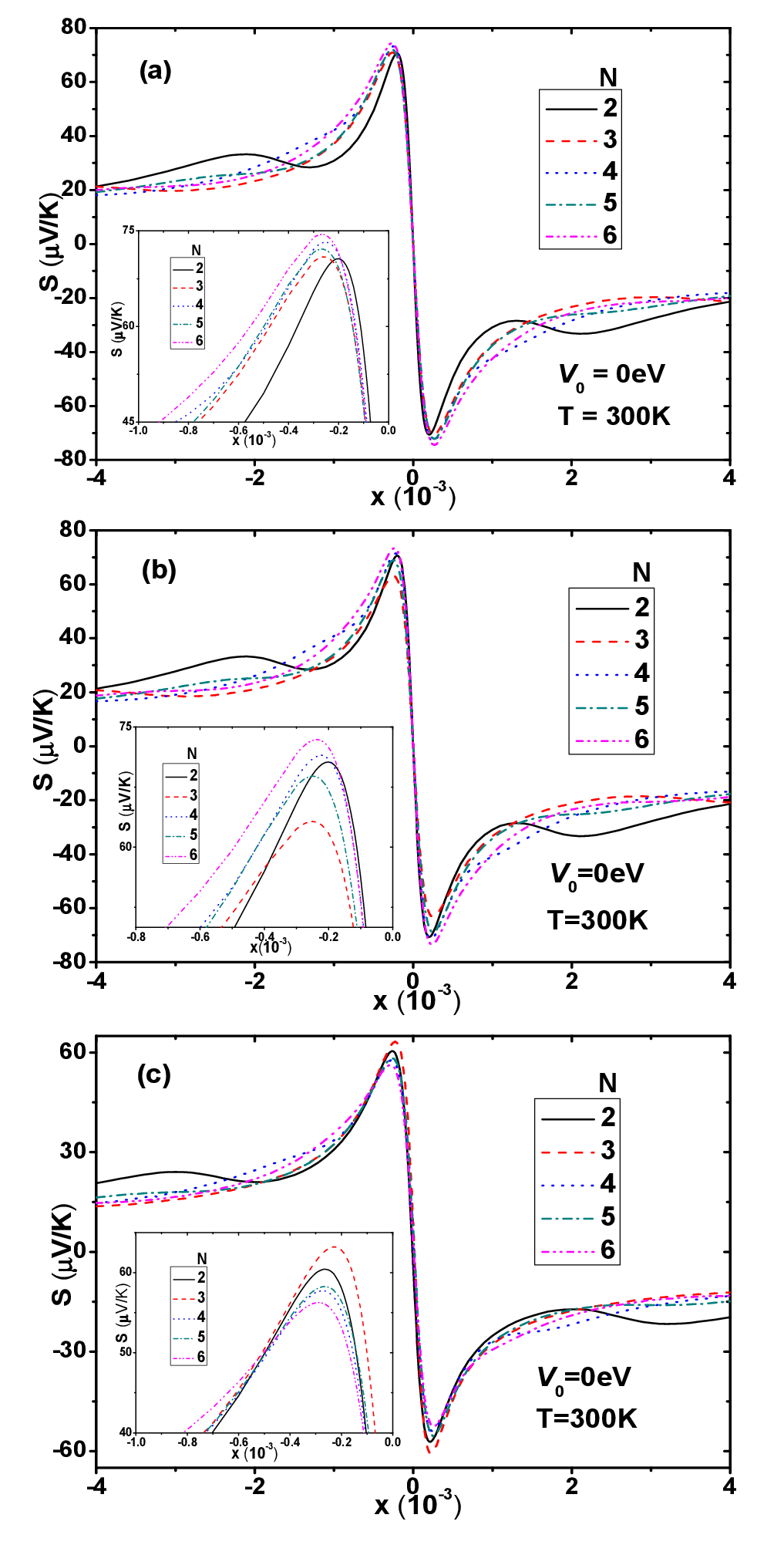}
\caption{Room temperature (300 K) thermopower of clean unbiased multilayer graphene.
(a)Nearest neighbor hopping model without including the screening effect or
charge redistribution between layers.
(b)Nearest neighbor hopping model with the screening effect included.
(c)Full SWMcC model with screening effect incorporated.}
\end{figure}

Results presented in Fig. 5(a) does not take the effect of
screening into account, so electrons on different layers feel
the same electrostatic potential. However, as discussed in Sec.
II C, multilayer graphene with layer number larger than two
should have different carrier densities for different layers.
When this effect is treated self-consistently, results for the
multilayer graphene described in terms of the simplified model
are presented in Fig. 5(b). Except for some quantitative
difference, the qualitative behavior is the same as in Fig.
5(a). The positions and magnitudes of various peaks are
separated into two groups depending on whether $N$ is  even  or
odd. For both cases, the peak value of thermopower increases
and the peak position continuously shifts to higher carrier
density with increasing layer number. In the presence of
screening effect, site energies of the various layers become
unequal, hence the subsystem decomposition is not valid and
Fig. 5(b) is obtained starting from the full model.

Fig. 5(c) shows the thermopower of multilayer graphene
described in terms of the more general
Slonczewski-Weiss-McClure
model\cite{slonczewski58,mcclure57,nilsson08,partoens06,partoens07,avetisyan09},
with the screening effect incorporated. Since the particle-hole
symmetry is now explicitly broken, the curve becomes asymmetric
with respect to the point of zero carrier density.
Specifically, the positive and negative peak value becomes
unequal and the position of zero thermopower is shifted away
from zero doping. Another qualitative difference as compared to
the results obtained from the simplified model is that, the
peak value of the even-layer ($N$=2, 4, 6) and odd-layer
($N$=3, 5) systems show opposite behaviors compared to Fig.
5(b). This nontrivial result arising from further neighbor
hoppings contained in the SWMcC model needs experimental
confirmation.

The reason for the above grouping between even and odd layer
systems could be understood qualitatively from the subsystem
decomposition in Sec. IIB. Thermopower of various bilayer like
graphene subsystems form a series which show monotonous
dependency on the effective interlayer hopping $\gamma_{1m}$ in
Eq. (\ref{subblg}). Thermopower of monolayer like subsystem
does not fall into this series. Since monolayer like subsystems
are contained only in odd-layer graphene, it makes sense that
the even layer and odd layer multilayer graphene should group
separately. Though self-consistent calculation and inclusion of
further neighbor hopping both make the subsystem decomposition
invalid, numerical results contained in Figs. 5(b) and 5(c) are
still consistent with the above picture.

\subsection{Biased multilayer graphene: The electronic structure}

Transport properties, like conductivity and thermopower,
sensitively depend on the underlying electronic structure. In
order to understand the thermopower of biased multilayer
graphene to be presented in the next section, we first discuss
in this section the electronic structure of a biased multilayer
graphene, taking trilayer and quad-layer graphene systems as
two examples. In the presence of an electric field
perpendicular to the layer plane, inversion symmetry of a
multilayer graphene is broken explicitly. In this case, even
for the bilayer system, charge redistributes to screen the
electric field. As in the above section, we take this screening
into account in terms of the Hartree approximation by treating
separate layers as parallel plates with certain density of net
charge which are to be determined
self-consistently.\cite{mccann06,guinea07,avetisyan09,koshino09a,koshino10,avetisyan092}

We focus on two related questions. One is the robustness of the
Dirac-like linear dispersive band in the presence of an
electric field for a multilayer graphene with an odd number of
layers. The other is the possibility of opening a full energy
gap by an external electric field, similar to the
BLG.\cite{mccann06,min07,hao10}

We find that the effect of temperature on the electronic
structure is very tiny. Thus, only zero temperature results are
presented here. For the biased systems, without loss of
generality, the bare interlayer potential energy difference
induced by the electric field is taken as $V_{0}$=1 eV.

\begin{figure}
\centering
\includegraphics[width=9cm,height=8cm,angle=0]{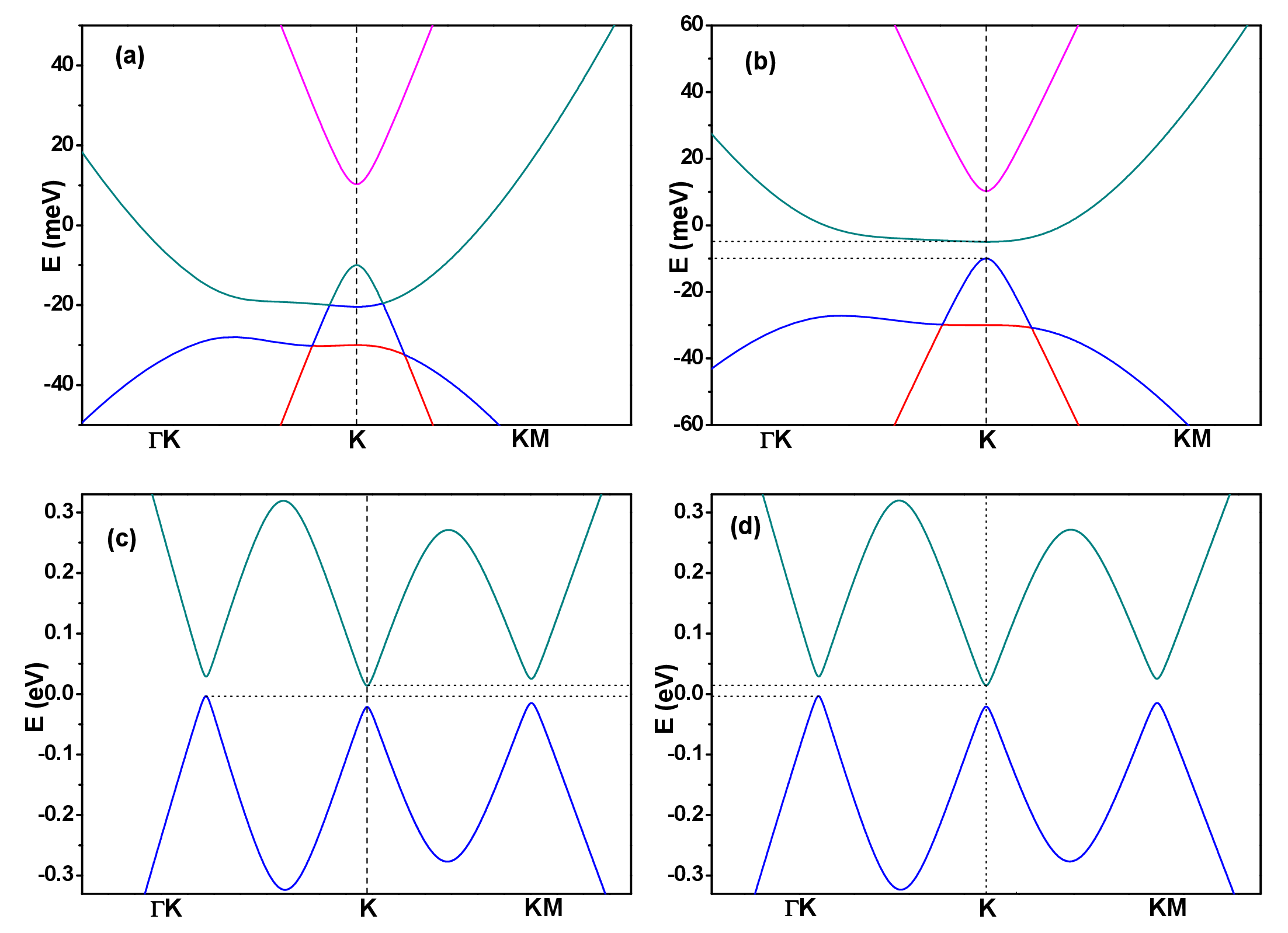}
\caption{Low energy band structures for unbiased and biased trilayer graphene.
(a) $V_{0}$=0, $n$=0, $\mu$$\simeq$-20.8 meV.
(b) $V_{0}$=0, $n$=2$\times$10$^{13}$ cm$^{-2}$, $\mu$$\simeq$542.1 meV.
(c) $V_{0}$=1 eV, $n$=0, $\mu$$\simeq$2.4 meV.
(d) $V_{0}$=1 eV, $n$=2$\times$10$^{12}$ cm$^{-2}$, $\mu$$\simeq$45.8 meV.
$n$ is the average carrier density per layer. $\Gamma$K and KM denote
respectively segments along the high symmetry lines in the two dimensional
BZ connecting two of three highly symmetric points $\Gamma$, K or M.}
\end{figure}

\begin{figure}
\centering
\includegraphics[width=9cm,height=8cm,angle=0]{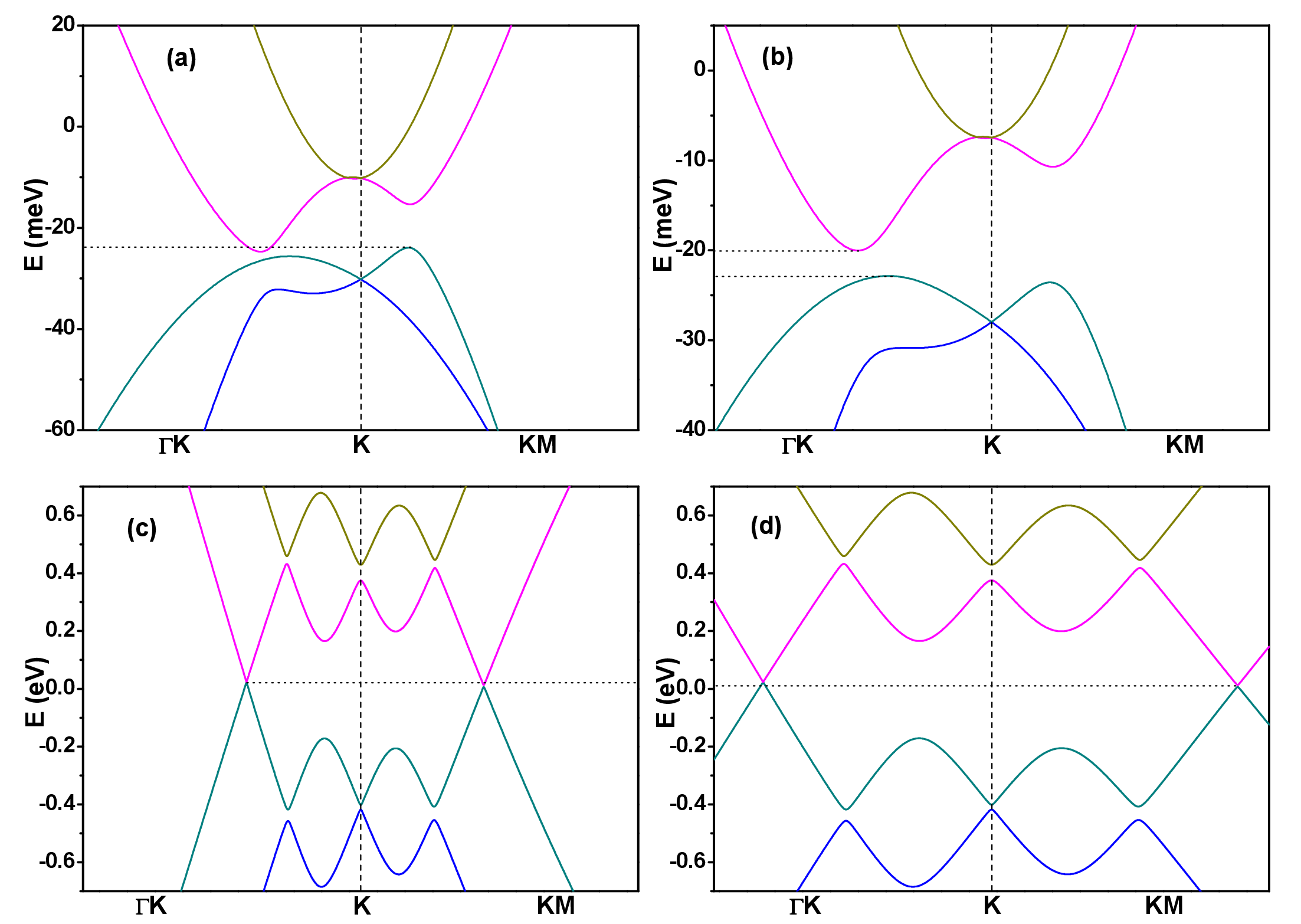}
\caption{Low energy band structures for unbiased and biased quad-layer graphene.
(a) $V_{0}$=0, $n$=0, $\mu$$\simeq$-20.8 meV.
(b) $V_{0}$=0, $n$=2.5$\times$10$^{12}$ cm$^{-2}$, $\mu$$\simeq$124.3 meV.
(c) $V_{0}$=1 eV, $n$=0, $\mu$$\simeq$15.5 meV.
(d) $V_{0}$=1 eV, $n$=2.5$\times$10$^{12}$ cm$^{-2}$, $\mu$$\simeq$43 meV.
Meanings of
$\Gamma$K and KM are the same as in Fig. 6.}
\end{figure}

For an unbiased charge neutral trilayer graphene, as shown in
Figs. 6(a), the system is gapless. But when the carrier density
is increased to be as high as 2$\times$10$^{13}$ cm$^{-2}$, a
small gap about 5 meV is induced below the chemical potential.
Two Dirac cone like structures still exist in the system, but
they do not touch at the cone vertex and are usually immersed
in bilayer like bands. Applying a perpendicular electric field
as high as $V_{0}$=1 eV, a small gap about $17$ meV is induced
even for the neutral system $x$=0, as shown in Figs. 6(c). The
magnitude of the gap enhances slightly with the increase of $x$
as shown in Fig. 6(d).

For the biased quad-layer graphene, as shown in Figs. 7(c) and
7(d), a full gap is not observed. For the unbiased quad-layer
graphene, though the neutral system is still gapless (Fig.
7(a)), a full gap below chemical potential is observed when the
carrier density is high enough, as shown in Fig. 7(b). The
latter feature, together with that in Fig. 6(b), arises from
the screening
effect.\cite{partoens06,partoens07,koshino07,avetisyan092}
Screening effect for the unbiased multilayer graphene systems
for $N$$>$2 matters because, for layer number larger than two,
$N/2$ (for even $N$) or ($N+1$)/2 (for odd $N$) non-equivalent
layer sets appear which usually have different on site energies
and electron densities.

\begin{figure}
\centering
\includegraphics[width=9cm,height=8cm,angle=0]{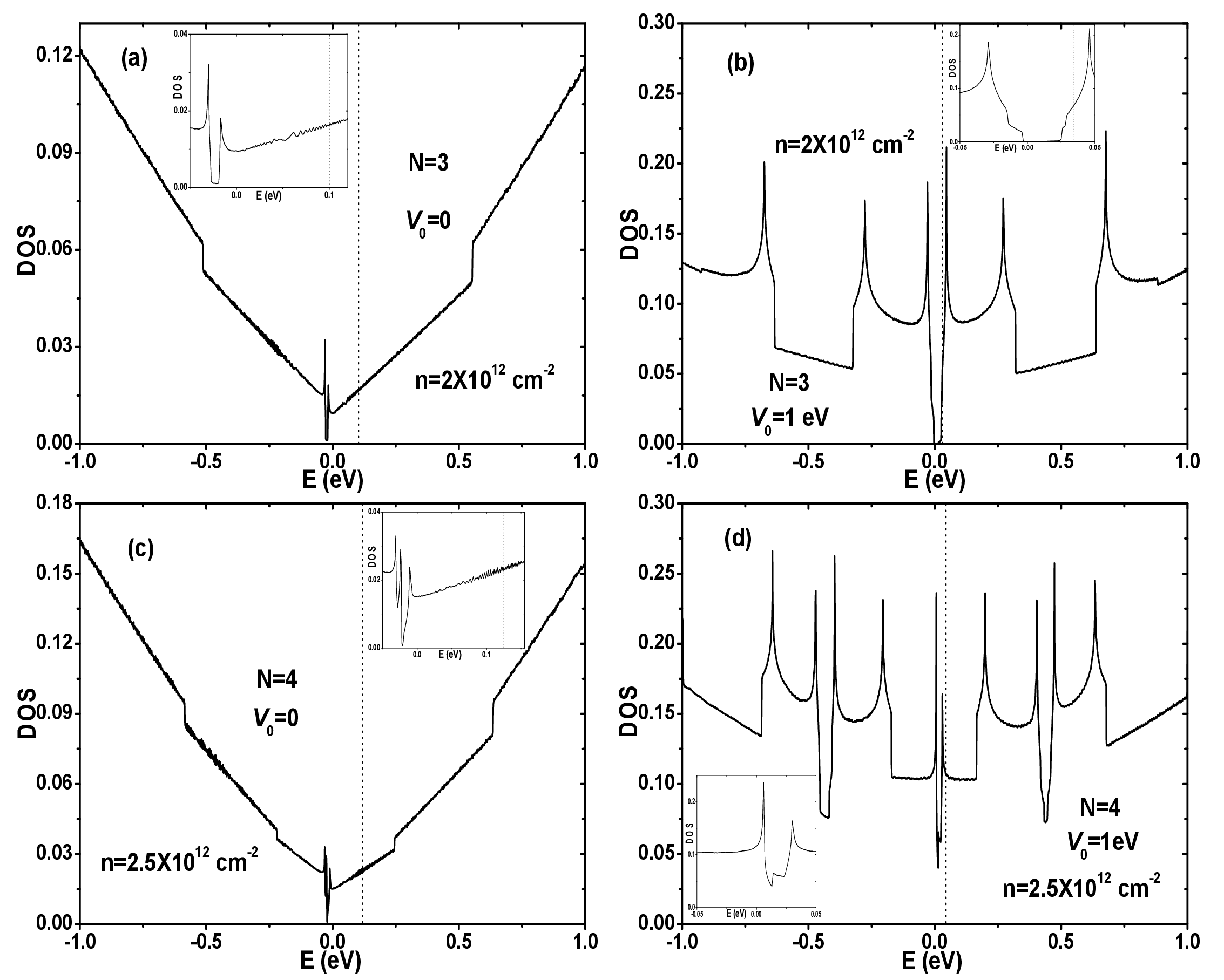}
\caption{Low energy density of states (DOS) for unbiased and biased trilayer and quad-layer graphene,
normalized to be corresponding to 6 (8) carbon atoms for trilayer (quad-layer) graphene per spin.
(a)Unbiased trilayer graphene, with $n$=2$\times$10$^{12}$ cm$^{-2}$. Room temperature chemical potential
is $\mu$(300K)$\simeq$ 101 meV.
(b)Biased trilayer graphene, with $n$=2$\times$10$^{12}$ cm$^{-2}$ and $V_{0}$=1 eV. $\mu$(300K)$\simeq$ 33.2 meV.
(c)Unbiased quad-layer graphene, with $n$=2.5$\times$10$^{12}$ cm$^{-2}$. $\mu$(300K)$\simeq$ 121 meV.
(d)Biased quad-layer graphene, with $n$=2.5$\times$10$^{12}$ cm$^{-2}$ and $V_{0}$=1 eV. $\mu$(300K)$\simeq$ 42.3 meV.
 Dotted vertical lines mark position of the
corresponding room temperature chemical potentials. Insets are enlargements of the small energy regions.}
\end{figure}

Finally, we present in Fig. 8 the density of states (DOS) of
the trilayer and quad-layer graphene for several typical
parameter sets. The corresponding room temperature chemical
potentials are marked by the dotted vertical lines. As compared
to the unbiased systems, external bias introduces some Van Hove
singularities in the density of states.

\subsection{Biased multilayer graphene: The thermopower}

Previously, we have shown that the opening of a gap in biased
BLG greatly enhances  thermopower of the
system\cite{hao10,kuroki07}. Calculations in the previous
section show that a small gap opens also in other multilayer
graphene systems, such as the trilayer graphene. It is thus
interesting to see whether or not  the thermopower is likewise
enhanced in these systems.

Results for multilayer graphene systems at room temperature
with layer number up to $5$  are presented in Fig. 9. Strength
of the external electric field is taken as $V_{0}$=1 eV. Fig.
9(a) contains results obtained from the nearest neighbor
hopping model. Fig. 9(b) is for the SWMcC model.

\begin{figure}[tbp]
\centering
\includegraphics[width=8cm,height=15cm,angle=0]{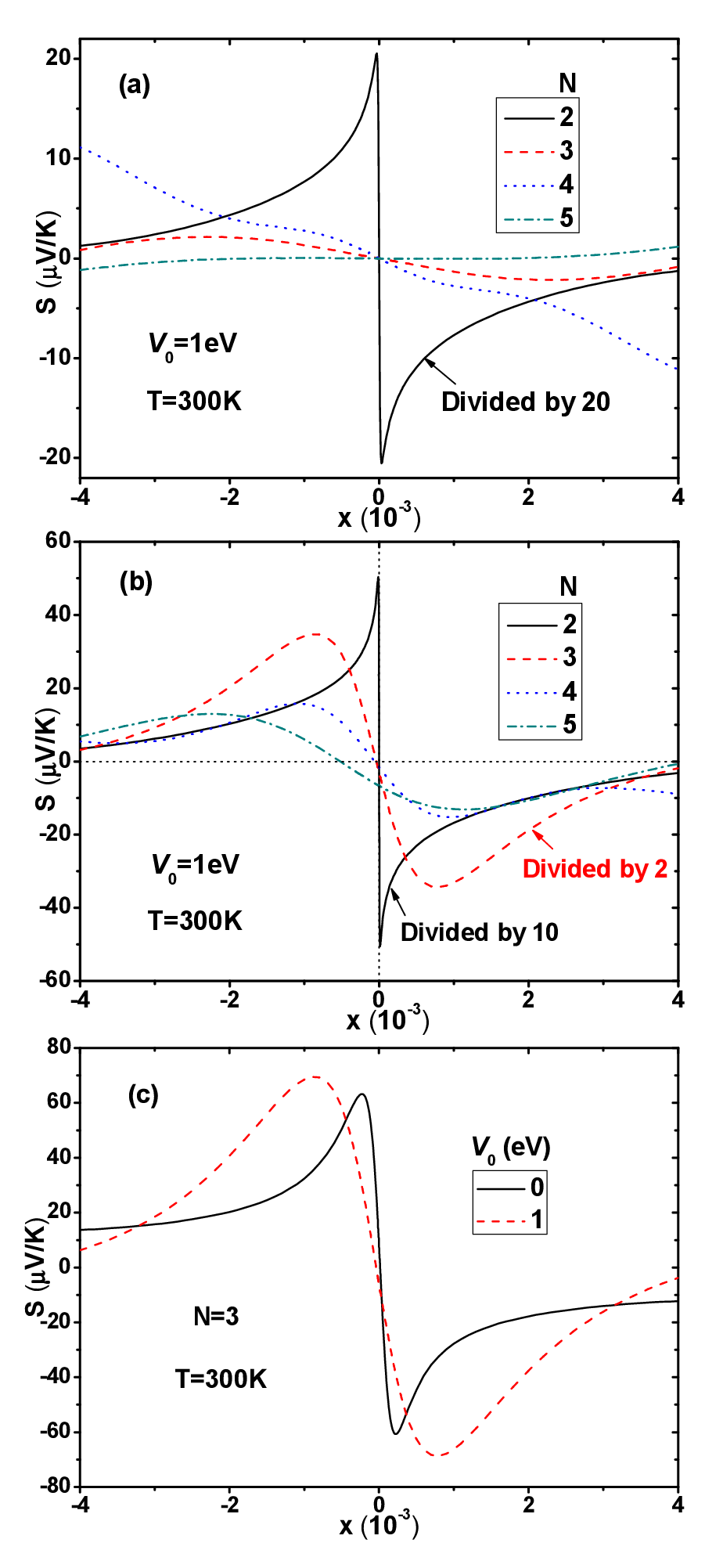}
\caption{Room temperature (300 K) thermopower of biased multilayer graphene
as a function of carrier density.
(a)Nearest neighbor hopping model with the screening effect taken into account.
(b)SWMcC model with screening effect incorporated. (c)A trilayer graphene
with and without a bias potential.}
\end{figure}

Compared with zero bias results, the large enhancement of
thermopower under a bias is specific to the BLG system. Though
peak value of thermopower is slightly enhanced in trilayer
graphene (Fig. 9(c)), it decreases under a bias for both the
quad-layer and quintuple-layer systems.

In a biased BLG, peak value of room temperature thermopower is
roughly proportional to the gap size. An external field as
large as $V_{0}$=1 eV opens a gap of about 288 meV, which
yields a peak room temperature thermopower of about 412
$\mu$V/K.\cite{hao10} From the previous subsection, $V_{0}$=1
eV gives a gap of about 20 meV for trilayer graphene. Following
this thread of argument, room temperature thermopower of the
biased ($V_{0}$=1 eV) trilayer graphene would be around 50
$\mu$V/K, which is in qualitative agreement with Fig. 9(b).

For layer number larger than three (e.g., Fig. 7(c) and 7(d)
for the quad-layer system), usually no full gap opens in the
biased multilayer graphene system since the bottom and top
layers almost decouple for so thick samples. In addition, as
could be seen from Figs. (6), (7) and (8), the chemical
potential is usually not situated inside the gap, even though a
gap opens for certain doping and external electric field. Both
of these effects suppress the thermopower for biased multilayer
graphene systems with $N$$\ge$3.

To confirm the self-consistency of the results, we analyze the
applicability of the Mott's formula, which relates the
thermopower to the longitudinal conductivity
as\cite{hwang09,lofwander07,yan09,cutler69}
\begin{equation}
S=-\frac{\pi^2 k_{B}^{2}T}{3e}\frac{\partial\text{ln}\sigma(\mu)}{\partial\mu}.
\end{equation}
The conductivity is obtained from the linear response
coefficient as $\sigma$=$e^2 L_{11}/T$. We present in Fig. 10
results obtained by a full microscopic calculation and by
fitting the Mott's formula, taking the room temperature
thermopower of biased ($V_{0}$=1 eV) trilayer and quad-layer
systems as two examples. The Mott's formula holds well at
temperatures smaller compared to the Fermi energy. As shown in
Fig. 10, the qualitative behavior of thermopower is reproduced
very well by the Mott's formula even at the room temperature.
Only close to the peak position, is the difference appreciable.

\begin{figure}[tbp]
\centering
\includegraphics[width=7cm,height=10cm,angle=0]{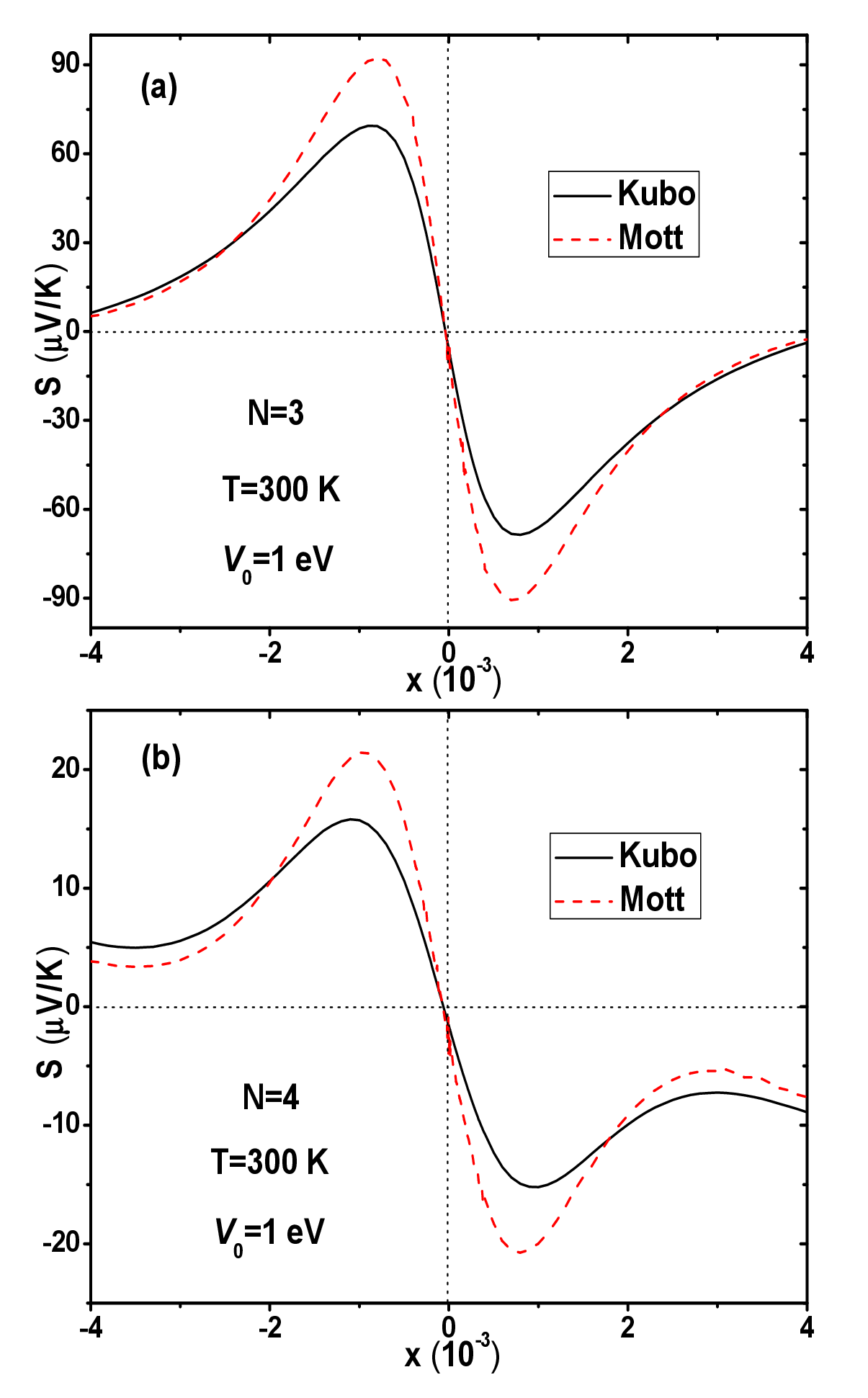}
\caption{Room temperature (300 K) thermopower of biased ($V_{0}$=1 eV)
(a) trilayer and (b) quad-layer graphene.
A comparison of results from full microscopic calculation (labeled as ``Kubo")
and a fitting from the Mott's formula (labeled as ``Mott") is made.
The SWMcC model including the screening effect is considered here.}
\end{figure}

From experiences gained on the monolayer and bilayer graphene,
it is safe to say that impurity scattering would not enhance
thermopower much. Thus, we conclude that biased BLG systems has
the largest room temperature thermopower among all the
multilayer graphene systems.

\section{summary}
In this work, we systematically calculate thermopower of biased
and unbiased multilayer grphene systems. The effect of
screening of an external electric field is taken into account
self-consistently under the Hartree approximation so that charge densities are different between inner and outer layers.
 Both the
model with only nearest neighbor hopping and the more general
SWMcC model with further neighbor hoppings are considered. The
effect of impurity scattering is considered for monolayer and
unbiased bilayer graphene in terms of the self-consistent Born
approximation. For monolayer graphene, only when the effect of
impurity scattering is taken into account, we could obtain
results consistent with experiments. The electronic structure
and thermopower of biased multilayer graphene systems are
calculated in which the screening effect is self-consistently
incorporated. The biased bilayer graphene shows the largest
room temperature thermopower.

\begin{acknowledgments}
We wish to acknowledge the support of NSC under Grant No.
98-2112-M-001-017-MY3.
\end{acknowledgments}\index{}


\end{document}